\documentclass[prb,superscriptaddress,amsmath,amssymb,floatfix,showpacs,twocolumn]{revtex4-1}

\usepackage{amsfonts,amsmath,amssymb}
\usepackage{times}
\usepackage{graphicx}
\usepackage{bm} 
\usepackage{color}
\usepackage{dcolumn} 
\usepackage{array}
\usepackage{float}
\newcommand{\up}{\uparrow}
\newcommand{\dn}{\downarrow}
\newcommand{\s}{\sigma}
\renewcommand{\sb}{\bar{\sigma}}
\newcommand{\rno}{RNiO$_3$}

\newcommand{\luno}{L\lowercase{u}N\lowercase{i}O$_3$}

\newcolumntype{d}{D{.}{.}{1.2}}
\newcommand{\LB}{\textsc{LB}}
\newcommand{\SB}{\textsc{SB}}
\newcommand{\DFT}{\textsc{DFT}}

\def\hn{\hat{n}}
\def\hH{\hat{H}}
\newcommand{\Peg}{P_{e_g}}

\newcommand{\Ueffuu}{U_{\mathrm{eff}}^{\s\s}}

\newcommand{\comment}[1]{{#1}}

\begin{document}

\title{Renormalization of effective interactions in a negative charge-transfer insulator}

\author{Priyanka Seth}
\affiliation{Centre de Physique Th\'eorique, \'Ecole Polytechnique, CNRS, Universit\'e Paris-Saclay, 91128 Palaiseau, France}
  \affiliation{Institut de Physique Th\'eorique (IPhT), CEA, CNRS, 91191 Gif-sur-Yvette, France}
\author{Oleg E. Peil}
  \affiliation{Centre de Physique Th\'eorique, \'Ecole Polytechnique, CNRS, Universit\'e Paris-Saclay, 91128 Palaiseau, France}
  \affiliation{Department of Quantum Matter Physics, University of Geneva, 24 
Quai Ernest-Ansermet, 1211 Geneva 4, Switzerland}
  \affiliation{Coll\`ege de France, 11 place Marcelin Berthelot, 75005 Paris, 
France}
\author{Leonid Pourovskii}
\affiliation{Centre de Physique Th\'eorique, \'Ecole Polytechnique, CNRS, Universit\'e Paris-Saclay, 91128 Palaiseau, France}
  \affiliation{Coll\`ege de France, 11 place Marcelin Berthelot, 75005 Paris, 
  	France}
\author{Markus Betzinger}
  \affiliation{Peter Gr\"unberg Institut and Institute for Advanced Simulation, Forschungszentrum J\"ulich and JARA, 52425 J\"ulich, Germany}
\author{Christoph Friedrich}
  \affiliation{Peter Gr\"unberg Institut and Institute for Advanced Simulation, Forschungszentrum J\"ulich and JARA, 52425 J\"ulich, Germany}
\author{Olivier Parcollet}
  \affiliation{Institut de Physique Th\'eorique (IPhT), CEA, CNRS, 91191 Gif-sur-Yvette, France}
\author{Silke Biermann}
\affiliation{Centre de Physique Th\'eorique, \'Ecole Polytechnique, CNRS, Universit\'e Paris-Saclay, 91128 Palaiseau, France}
\author{Ferdi Aryasetiawan}
  \affiliation{Department of Physics, Division of Mathematical Physics, Professorsgatan 1, 223 62 Lund, Sweden}
\author{Antoine Georges} 
\affiliation{Centre de Physique Th\'eorique, \'Ecole Polytechnique, CNRS, Universit\'e Paris-Saclay, 91128 Palaiseau, France}
  \affiliation{Department of Quantum Matter Physics, University of Geneva, 24 
Quai Ernest-Ansermet, 1211 Geneva 4, Switzerland}
  \affiliation{Coll\`ege de France, 11 place Marcelin Berthelot, 75005 Paris, 
France}

\date{\today}

\begin{abstract}
We compute from first principles the effective interaction parameters appropriate for a 
low-energy description of the rare-earth nickelate \luno{} involving the partially occupied 
$e_g$ states only. 
The calculation uses the constrained random-phase approximation and reveals 
that the effective on-site Coulomb repulsion is strongly reduced by screening effects 
involving the oxygen-$p$ and nickel-$t_{2g}$ states.
The long-range component of the effective low-energy interaction is also 
found to be sizeable. 
As a result, the effective on-site interaction between parallel-spin electrons 
is reduced down to a small negative value. This validates effective low-energy 
theories of these materials proposed earlier. 
Electronic structure methods combined with dynamical mean-field theory are used 
to construct and solve an appropriate low-energy model and explore its phase 
diagram as a function of the on-site repulsion and Hund's coupling. 
For the calculated values of these effective interactions we find, in agreement with experiments, 
that \luno{} is a metal without disproportionation of the $e_g$ occupancy when  
considered in its orthorhombic structure, while the monoclinic phase 
is a disproportionated insulator. 

\end{abstract}


\maketitle

\newpage

\section{Introduction}

The interplay between the atomic physics and strong covalent bonding in
transition-metal oxides (TMO) results in a variety of fascinating phenomena
\cite{Imada1998}. 
The energy scale spanned by the hybridized states
formed by the $d$ orbitals of the transition metal and the $p$ states of oxygen 
is typically of order $10$~eV. However, it is often useful for physical understanding to use 
a ``low-energy'' description in which only a subset of the metal-oxygen antibonding states 
is retained, namely the partially occupied states in proximity to the Fermi level. 
Those usually span a narrower energy window of a few electron-Volts. 
Going over from the full high-energy description to a low-energy model allows
one to reduce the dimension of the Hilbert space considerably, and quite often
provides physical insight into the behavior of a material. This is particularly relevant 
to late transition-metal oxides involving antibonding $e_g$ orbitals, as exemplified 
by the Zhang-Rice single-band picture of cuprates \cite{Zhang1988}. 

A price to pay for this simplification is the renormalization of interaction parameters 
when high energy states are integrated out. These renormalizations can be large, 
and evaluating the proper values of low-energy interactions is a challenging problem 
of great practical importance. 
In all TMOs, an important interaction is the Coulomb repulsion $U_{dd}$
between localized, atomic-like, $d$ states of the TM cation.  In late TMOs,
however, the energy scale relevant for low-energy states is the charge-transfer
energy, which can be much smaller\cite{Zaanen1985} than $U_{dd}$. 

A class of materials in which this issue is particularly relevant is the family of 
rare-earth nickelates, \rno{}. These materials have a very large degree of 
covalency between the Ni and O states\cite{varignon_2016}. This may result in the charge-transfer 
energy being very small in magnitude and 
possibly negative\cite{Mizokawa1991,Mizokawa1994,Mizokawa2000,Johnston2014,Strand2014}, 
leading to the appearance of holes on ligand (oxygen) states in the ground-state\cite{Demourgues1993}. 
A direct confirmation of the presence of ligand holes has been recently provided by 
X-ray absorption and resonant inelastic X-ray scattering experiments\cite{bisogni_2016}.  

The metal-insulator transition (MIT) of the \rno{} series is accompanied by a structural transition from
the high-$T$ orthorhombic structure to a low-$T$ monoclinic structure. In the latter, 
the uniform octahedra of the orthorhombic structure distort into a set of
compressed octahedra with short Ni-O bonds (SB) and a set of expanded octahedra with
long bonds (LB). A qualitative, somewhat extreme, picture of the low-$T$ phase\cite{Park2012,Johnston2014} 
is to assign the configuration $d^8$ to the Ni sites of the LB octahedra and $d^8\overline{L}^2$ 
(with two ligand holes delocalized on neighboring oxygens) to the SB octahedra. This is 
in contrast to the nominal valence Ni-$d^7$ suggested by a naive counting in the ionic limit 
(with $R^{3+}$, $O^{2-}$) so that the picture above can be summarized as $d^7+d^7 \rightarrow d^8+d^8\overline{L}^2$.  
Correspondingly, in this extreme picture, the LB sites would carry a spin-$1$ magnetic moment, while 
the SB sites would carry no magnetic moment 
(the Ni-moment being screened by the oxygen holes\cite{Park2012}). 
Note that this disproportionation does not necessarily correspond to a large amplitude charge ordering 
since each oxygen actually belongs to both a SB and LB octahedron so that the average 
charge on each octahedron can remain weakly modulated or even uniform. 

\begin{figure}[htpb]
  \centering
    \includegraphics[width=1.00\columnwidth]{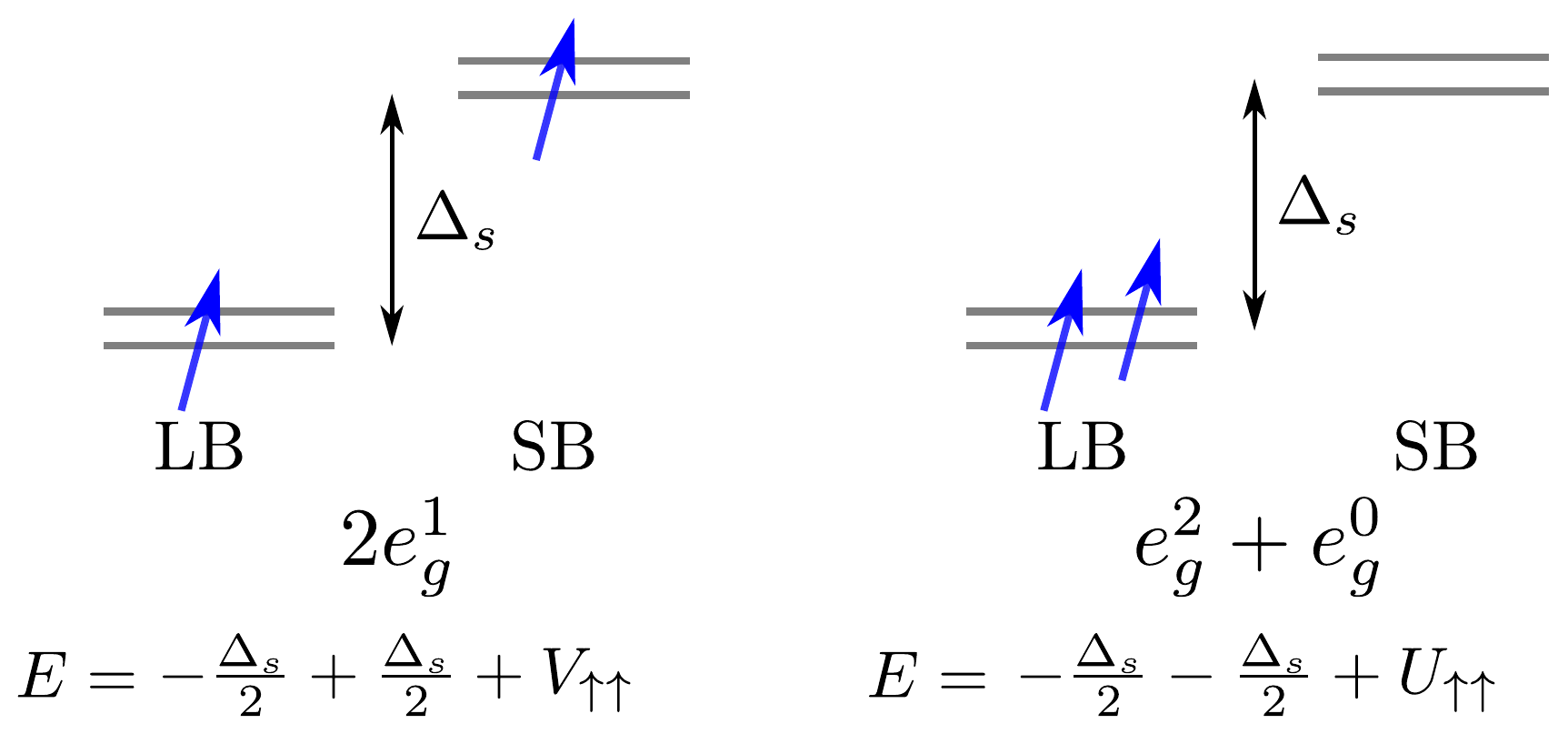}
    \caption{Schematic picture of the disproportionation associated with the metal-insulator 
    transition for a two-site model consisting of a LB site and a SB site. 
    The energy of the LB site is lowered as compared to that of the SB site by 
    the Peierls energy $\Delta_s$. Each site carries two $e_g$ orbitals. In the doubly occupied 
    configuration, each orbital is occupied by a single electron with parallel spins, in accordance with 
    Hund's rules. An on-site interaction $U^{\s\s}$ between electrons with parallel spins is considered, 
    as well as an inter-site interaction $V^{\s\s}$. In the atomic limit where hopping is neglected, 
    the criterion for stabilizing the disproportionated state $e_g^2+e_g^0$ over the uniform one 
    $e_g^1+e_g^1$ reads: $U^{\s\s}-V^{\s\s}<\Delta_s$. The extension of this criterion to a whole 
    lattice in the presence of hopping is discussed in the main text. 
    \label{fig:disproportionation}}
\end{figure}

Subedi \textit{et al.} \cite{subedi_low-energy_2015} recently proposed a low-energy description 
of the electronic structure of the \rno{} series, involving only the $p-d$ hybridized 
antibonding states with $e_g$ symmetry close to the Fermi level. In this description, 
the above disproportionation can be viewed as $e_g^1+e_g^1\rightarrow e_g^2+e_g^0$ 
(Fig.~\ref{fig:disproportionation}).  
Building on earlier ideas by Mazin et al.\cite{mazin_charge_2007}, these authors showed 
that such a disproportionation is favored by a strong reduction of the effective $U$ 
acting on the low-energy $e_g$ states and by a large value of the low-energy Hund's 
coupling $J$.  
More precisely, the monoclinic distortion splits the low-energy $e_g$ states into 
two groups of states separated by a Peierls-like energy gap $\Delta_s$ (note that 
this gap opens at the energy corresponding to half-filling, and is hence not 
directly responsible for the transition into the insulating state of these nominally 
quarter-filled compounds). Using dynamical mean-field theory (DMFT)\cite{georges_dynamical_1996} 
in combination with density-functional calculations, a $U$-$J$ phase diagram was established 
for the low-energy model, demonstrating that 
a disproportionated insulating phase is present in the range of coupling parameters where the 
parallel-spin interaction $U^{\s\s} = U-3J$ is smaller than the Peierls gap $\Delta_s$. 

The low-energy picture of Subedi {\it et al.}\cite{subedi_low-energy_2015} is in good agreement
both with experiments (e.g. optical spectroscopy\cite{ruppen_optical_2015,ruppen_2017}) 
and with earlier DMFT calculations including all Ni-$d$ and O-$p$ states\cite{Park2012}. 
However, the question of whether the strongly renormalized value of the effective 
low-energy interaction $U^{\s\s}=U-3J$ is indeed realistic remains widely open. 
A first-principles calculation of these low-energy effective interactions is obviously highly 
desirable. 
Furthermore, Ref.\onlinecite{subedi_low-energy_2015} did not consider the role 
of inter-site interactions, which are surely induced when downfolding onto a low-energy model 
and are known to be important in materials with electronic disproportionation or charge ordering.
When the intersite interaction $V$ between the LB and SB sites is included, the more accurate condition 
for charge disproportionation becomes $U_{\s \s} - V_{\s \s} < \Delta_s$ for the two-site case treated in the atomic limit, 
as depicted in Fig.~\ref{fig:disproportionation}. For the full problem one needs to include hopping, correctly treat the lattice connectivity, and also include the effect of long-range Coulomb interactions, which are important in insulators. 
There is little experimental spectroscopic information on 
such long-range interactions.
The aim of the present article is to attempt a first-principles determination of the appropriate 
low-energy parameters, and examine the physical consequences of the obtained values in light of the 
issues discussed above.  

The approach that we shall adopt is the constrained random-phase approximation (cRPA) 
\cite{aryasetiawan_frequency-dependent_2004}. 
This method has proven successful in  
calculating interaction parameters between electrons in localized
$d$ or $f$ states assumed to be screened by more extended $s$ and $p$ states
\cite{PhysRevB.77.085122, VaugierPRB86-2012, PhysRevB.85.045132, PhysRevB.89.125110, PhysRevB.94.125147, 
NiO-U, AcO-U}. 
In this paper, we apply this method to calculate the interaction parameters
corresponding to low-energy states of \luno{}, which exhibits the largest
distortion amongst the family of \rno{}. 
It is worth emphasizing that such a system represents a true challenge to cRPA because 
the contributions to screening come both from extended O-$p$, with possible ligand holes, 
which are very close in energy to the $e_g$ states and 
strongly hybridized, and also from localized completely filled $t_{2g}$ states of the Ni ions. 

In this paper we show that, despite these challenges, 
the cRPA method is indeed able to produce the large renormalization of the Coulomb repulsion $U$. 
We also show that $U_{\s \s}$ is further effectively reduced due to intersite Coulomb interactions 
down to values comparable to the Peierls gap $\Delta_s$, hence establishing on firm grounds  
the low-energy description suggested previously\cite{mazin_charge_2007,subedi_low-energy_2015}, 
with the additional twist of large non-local interactions effectively renormalizing the local ones.
We calculate the phase diagram of \luno{} within a combination of density-functional theory-based 
electronic structure and dynamical mean-field theory (DFT+DMFT), including the intersite
interactions at a static mean-field level. 
For the cRPA values of $U$, $J$ and of the inter-site interactions, our DFT+DMFT 
calculations yield a metallic state for the orthorhombic phase and a disproportionated insulator 
for the monoclinic one, in agreement with experiments. 

This article is organized as follows.
In Sec.~\ref{sec:low_e_model}, we provide an introduction to the electronic structure of 
\luno{} and to the effective low-energy description in terms of $e_g$ states. 
In Sec.~\ref{sec:crpa} we implement the constrained random-phase approximation
and compute the resulting {\it ab initio} interaction parameters. 
In Sec.~\ref{sec:longrange_dmft}, we summarize the {\it ab initio} construction 
of the low-energy effective model and explore its phase diagram within the 
DFT+DMFT framework for both the orthorhombic and monoclinic phases, as 
a function of $U$ and $J$. We show that the cRPA-calculated values of these 
parameters correspond to a location of each of the two structures in 
this phase diagram which is physically consistent.   
Our results and findings are briefly summarized and discussed in Sec.~\ref{sec:discussion}.

%
%
\section{Electronic structure and low-energy model}
\label{sec:low_e_model}

\begin{figure}[htpb]
  \centering
    \includegraphics[width=1.00\columnwidth]{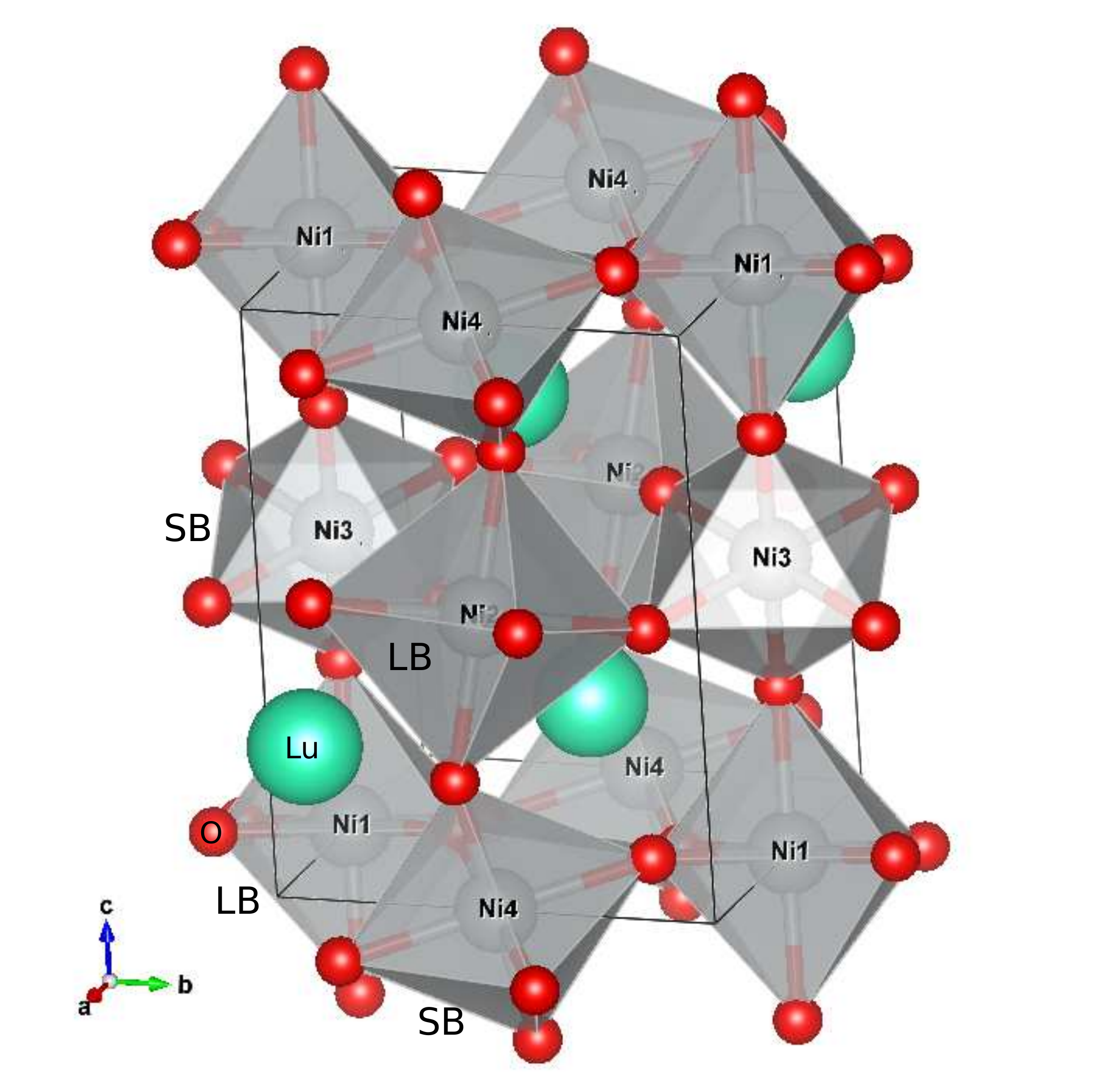}
    \caption{\luno{} in the monoclinic phase. Oxygen atoms (red) form distorted 
             octahedra containing Ni (grey). This structure is intercalated with Lu atoms
             (turquoise).  The short-bond (SB) and long-bond (LB) octahedra are identified.
             The labels of the Ni atoms correspond to the positions given in Table
             ~\ref{tab:ni_pos}.  
             \label{fig:luno_structure}}
\end{figure}

The electronic structure of both the low-temperature monoclinic (space group
$P2_1/n$, see Fig.~\ref{fig:luno_structure}) and high-temperature orthorhombic
($Pbnm$) phases of \luno{} have been calculated using the experimental lattice
structures provided in Ref.~\onlinecite{Alonso2000} (at $T = 673$K for $Pbnm$ and
$T = 533$K for $P2_1/n$). The unit cells of both
structures contain four formula units, but the monoclinic one differs by having
two distinct types of NiO$_6$ octahedra, one with short Ni-O bonds and one with
long bonds corresponding to compressed and expanded octahedra. 
For the reader's convenience we list the fractional coordinates of the Ni sites in the
monoclinic cell in Table~\ref{tab:ni_pos}. 

\begin{table}[htbp]
\caption{Fractional coordinates and types of Ni sites in the monoclinic
structure of \luno{}. Ni$_1$ and Ni$_2$ are of the LB type and lie diagonally across from each other, 
as do the SB sites Ni$_3$ and 
Ni$_4$. Ni$_1$ has as nearest neighbors 2 $\times$ Ni$_3$ in the $z$
direction and 4 $\times$ Ni$_4$ in the $x$-$y$ plane.}
\label{tab:ni_pos}
\centering
\begin{tabular}{lcccc}
       & type    & $X$       & $Y$           & $Z$           \\
\hline
Ni$_1$ & LB & $\frac{1}{2}$ & $0$           & $0$           \\
Ni$_2$ & LB & $0$           & $\frac{1}{2}$ & $\frac{1}{2}$ \\
Ni$_3$ & SB & $\frac{1}{2}$ & $0$           & $\frac{1}{2}$ \\
Ni$_4$ & SB & $0$           & $\frac{1}{2}$ & $0$           
\end{tabular}
\end{table}

In our density-functional-theory (DFT) calculations within the
local-density approximation (LDA) we have employed the full-potential augmented-plain-wave
(FLAPW) method as implemented in the \textsc{fleur} package
\cite{fleur_web,fleur}. All calculations were performed using a $\mathbf{k}$-mesh
consisting of $4 \times 4 \times 2$ points.

\begin{figure}[htpb]
  \centering
    \includegraphics[width=1.00\columnwidth]{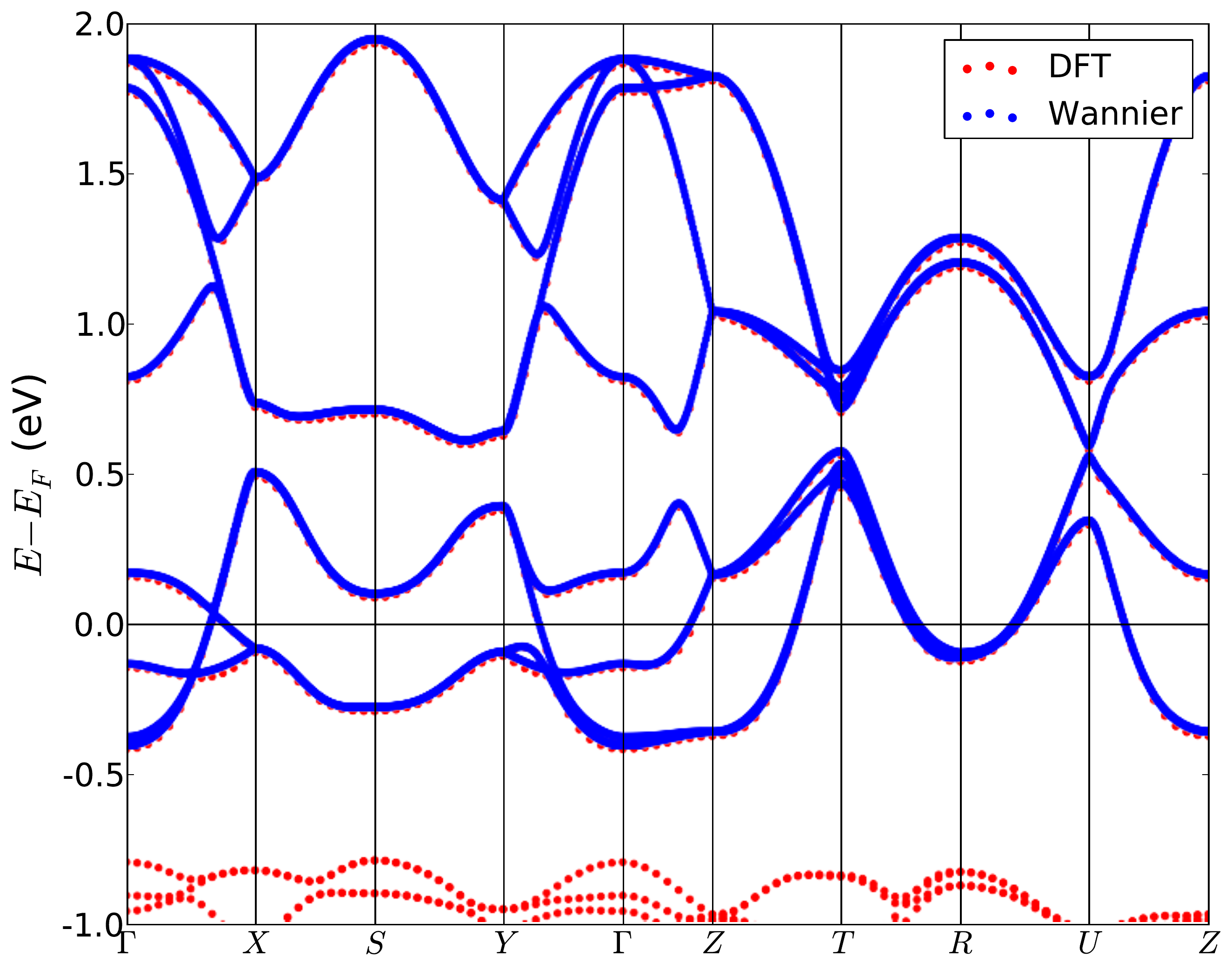}
    \caption{The band structure of the monoclinic phase of \luno{} for a unit
             cell consisting of four formula units. The DFT bands are shown in red and the
             band structure produced by the Wannier-basis low-energy Hamiltonian 
             is shown in blue. Within the numerical accuracy, 
             the Wannier bands are identical to the DFT $e_g$ bands. The 
             Peierls gap splits the set of 8 $e_g$ bands into two separated sets of 4 bands. 
             \label{fig:bands}}
\end{figure}

The calculated low-energy band structure of monoclinic \luno{}
(Fig.~\ref{fig:bands}) features a manifold of 8 $e_g$ bands in the range of
[-0.4:1.9] eV  around the Fermi level, with the filled $t_{2g}$ bands located
below -0.7~eV in energy and, hence, well separated from the $e_g$ ones. 
%
We note that a small `Peierls gap' $\Delta_{s}^{\mathrm{DFT}}\simeq 0.25$~eV  
separates the $e_g$ bands into 2 sets of 4 bands (except an isolated point $U$),
at an energy corresponding to 
the nominal filling of two electrons per site (half-filling), i.e. about $+0.5$~eV above the LDA Fermi level.
The Peierls gap originates from the existence of two types of sites in the distorted monoclinic phase: 
LB sites are pushed down in energy relative to the more
covalent SB sites, for which the Ni $d$ and O $p$ orbitals overlap more.
The DFT band structure of the orthorhombic phase is quite similar to that of the
monoclinic phase (see e.g. Fig.~1 of Ref.~\onlinecite{subedi_low-energy_2015}),
apart from the fact that the Peierls gap is of course closed in this case. 

In order to construct the quadratic part of the low-energy model, we downfold
the states around the Fermi energy. A set of maximally localized Wannier
functions representing correlated $e_g$ states on Ni sites was constructed from
the 8 $e_g$ bands using the \textsc{wannier90} package\cite{wannier90}. Given
the absence of entanglement between the $e_g$ bands and the rest of the electronic
structure in \luno{}, the projection from the space of Kohn-Sham
$e_g$ eigenstates onto the Wannier basis is simply a unitary transformation.
Hence, by diagonalizing the resulting low-energy Wannier Hamiltonian, one
reproduces the original DFT bands, as shown in Fig.~\ref{fig:bands}.

%
%
\section{Constrained-random-phase approximation}
\label{sec:crpa}

\subsection{Method description}

With the quadratic part of the low-energy Hamiltonian for the $e_g$
states, we perform {\it ab initio} calculations of the corresponding
interaction terms using the constrained random-phase approximation 
(cRPA)~\cite{aryasetiawan_frequency-dependent_2004}.
 
The main idea behind the cRPA is to write the effective interactions suitable for
a low-energy multi-band model as the matrix elements of a partially screened
interaction in the chosen localized basis.  Calculations within the cRPA method
start from the evaluation of the bare (unscreened) Coulomb vertex $v$.
The partial polarization function $P_r(\omega)$ is then calculated. 
This describes the screening of $v$ within RPA by including all particle-hole transitions 
except those within the low-energy $e_g$ subspace, i.e., $P_r(\omega) = P(\omega) - \Peg(\omega)$, where
$P(\omega) $ is the total polarization function and $\Peg(\omega)$ is the
contribution of all transitions within the low-energy subspace.
Processes within the low-energy subspace will be subsequently treated
explicitly by solving the low-energy Hamiltonian with more sophisticated
many-body techniques beyond the RPA, such as DMFT.  

The partially screened frequency-dependent interaction is obtained as
$W_r(\omega) = v/\left[1-P_r(\omega) v\right]$. 
The Hubbard interactions are then obtained from matrix elements
of the static limit $W_r(\omega=0)$ of this effective interaction.
An important advantage of the cRPA is its ability to easily treat intersite 
interactions, as demonstrated in Refs.~\onlinecite{0953-8984-25-9-094005,
PhysRevLett.110.166401, Hansmann_uncertainty}. 

The cRPA procedure is unambiguously defined when a subset of correlated states
is separated from the rest of the bands, as it is the case with the $e_{g}$ bands
of \luno{}. However, in the case of an entanglement
between correlated and uncorrelated bands one faces the problem of determining
which screened processes should be included in $P_r(\omega)$. The two schemes
that have been proposed to date to handle this issue, disentanglement
\cite{Miyake2009} and projection \cite{Sasioglu2011}, both give identical
results in our case with $\Peg(\omega)$ including all transitions within the
$e_g$ manifold, as the 8 $e_g$ bands are well separated from all other bands.

In our calculations, the on-site and inter-site effective interactions between  
Ni $e_g$ states are obtained using the cRPA functionality of  
the \textsc{spex} code, a GW code based on the \textsc{fleur}~\cite{fleur_web} electronic structure 
package~\footnote{We mention in passing that we have developed an application of the \textsc{TRIQS/DFTTools} 
package~\cite{Aichhorn2009, parcollet_triqs_2015, aichhorn_dfttools_2016,triqs,Michalicek2013} 
which provides an interface to the \textsc{FLEUR} electronic structure code}.  
800 Kohn-Sham bands up to the energy
cut-off of 140  eV were included in $P(\omega)$ and, correspondingly,
$P_r(\omega)$. In order to describe correctly semicore and high-energy
Kohn-Sham states we extended the FLAPW basis by including additional local
orbitals\cite{Friedrich2006,Michalicek2013}. Namely, we included local orbitals for the 3$s$,
4$s$, 5$s$, 3$p$, 4$p$, 5$p$, 3$d$, 4$d$, 4$f$ and 5$f$ shells of Ni;  2$s$,
3$s$, 2$p$, 3$p$, 3$d$, 4$d$, 4$f$ and 5$f$ shells of O; 5$s$, 6$s$, 7$s$,
5$p$, 6$p$, 7$p$, 5$d$, 6$d$, 4$f$,  and 5$f$ shells of Lu, as well as 6$g$ and
7$h$ shells on all atoms.

\subsection{cRPA Results for LuNiO$_3$}
\label{sec:crpa_results}

Here we discuss the interaction parameters for the low-energy models as
obtained by cRPA. We consider only the density-density interaction terms for
parallel and anti-parallel spins. Note that the coupling constants $J_X$ and $J_P$, corresponding respectively 
to the spin-flip and pair-hopping terms can be obtained from the density-density terms
under the assumption of rotational invariance.
All results are given in the basis of the eight 
$e_g$ orbitals, ordered as $z^2$ and $x^2-y^2$ for each of the four Ni sites as
given above. 
The results for the orthorhombic phase are given in
Table~\ref{tab:u_orthorhombic} and those for the monoclinic phase in
Table~\ref{tab:u_monoclinic}.

\begin{table*}[htbp]
\caption{$U_{mm'}$ for anti-parallel and parallel spins for orthorhombic \luno{}.
         All values in eV. Two-index parameters are related to four-index ones as
         $U_{mm'}^{\s\sb} = U_{mm'mm'}$, 
         $U_{mm'}^{\s\s} = U_{mm'mm'} - U_{mm'm'm}$.}
\label{tab:u_orthorhombic}
\centering
\begin{tabular}{@{\extracolsep{15pt}}dddddddd@{}}
   \multicolumn{8}{c}{$U_{mm'}^{\s\sb}$ (anti-parallel spin)} \\
   \multicolumn{1}{c}{Ni$_1$, $z^2$} & \multicolumn{1}{c}{Ni$_1$, $x^2-y^2$} &
   \multicolumn{1}{c}{Ni$_2$, $z^2$} & \multicolumn{1}{c}{Ni$_2$, $x^2-y^2$} &
   \multicolumn{1}{c}{Ni$_3$, $z^2$} & \multicolumn{1}{c}{Ni$_3$, $x^2-y^2$} &
   \multicolumn{1}{c}{Ni$_4$, $z^2$} & \multicolumn{1}{c}{Ni$_4$, $x^2-y^2$} \\
   1.77       & 0.99           & 0.32       & 0.30           & 0.62       & 0.43           & 0.39       & 0.42           \\
   0.99       & 1.54           & 0.30       & 0.27           & 0.43       & 0.33           & 0.40       & 0.45           \\
   0.32       & 0.30           & 1.77       & 0.99           & 0.39       & 0.40           & 0.62       & 0.43           \\
   0.30       & 0.27           & 0.99       & 1.54           & 0.42       & 0.45           & 0.43       & 0.33           \\
   0.62       & 0.43           & 0.39       & 0.42           & 1.77       & 0.99           & 0.32       & 0.30           \\
   0.43       & 0.33           & 0.40       & 0.45           & 0.99       & 1.54           & 0.30       & 0.27           \\
   0.39       & 0.40           & 0.62       & 0.43           & 0.32       & 0.30           & 1.77       & 0.99           \\
   0.42       & 0.45           & 0.43       & 0.33           & 0.30       & 0.27           & 0.99       & 1.54           \\
   \hline \\
   \multicolumn{8}{c}{$U_{mm'}^{\s\s}$ (parallel spin)} \\
   \multicolumn{1}{l}{Ni$_1$, $z^2$} & \multicolumn{1}{l}{Ni$_1$, $x^2-y^2$} &
   \multicolumn{1}{l}{Ni$_2$, $z^2$} & \multicolumn{1}{l}{Ni$_2$, $x^2-y^2$} &
   \multicolumn{1}{l}{Ni$_3$, $z^2$} & \multicolumn{1}{l}{Ni$_3$, $x^2-y^2$} &
   \multicolumn{1}{l}{Ni$_4$, $z^2$} & \multicolumn{1}{l}{Ni$_4$, $x^2-y^2$} \\
   0.00       & 0.66           & 0.32       & 0.30           & 0.59       & 0.42           & 0.38       & 0.41           \\
   0.66       & 0.00           & 0.30       & 0.27           & 0.42       & 0.33           & 0.39       & 0.42           \\
   0.32       & 0.30           & 0.00       & 0.66           & 0.38       & 0.39           & 0.59       & 0.42           \\
   0.30       & 0.27           & 0.66       & 0.00           & 0.41       & 0.42           & 0.42       & 0.33           \\
   0.59       & 0.42           & 0.38       & 0.41           & 0.00       & 0.66           & 0.32       & 0.30           \\
   0.42       & 0.33           & 0.39       & 0.42           & 0.66       & 0.00           & 0.30       & 0.27           \\
   0.38       & 0.39           & 0.59       & 0.42           & 0.32       & 0.30           & 0.00       & 0.66           \\
   0.41       & 0.42           & 0.42       & 0.33           & 0.30       & 0.27           & 0.66       & 0.00           \\
\end{tabular}
\end{table*}

\begin{table*}[htbp]
\caption{$U_{mm'}$ for anti-parallel and parallel spins for monoclinic \luno{}.
         All values in eV. Two-index parameters are related to four-index ones as
         $U_{mm'}^{\s\sb} = U_{mm'mm'}$, 
         $U_{mm'}^{\s\s} = U_{mm'mm'} - U_{mm'm'm}$.}
\label{tab:u_monoclinic}
\centering
\begin{tabular}{@{\extracolsep{15pt}}dddddddd@{}}
   \multicolumn{8}{c}{$U_{mm'}^{\s\sb}$ (anti-parallel spin)} \\
   \multicolumn{1}{c}{Ni$_1$, $z^2$} & \multicolumn{1}{c}{Ni$_1$, $x^2-y^2$} &
   \multicolumn{1}{c}{Ni$_2$, $z^2$} & \multicolumn{1}{c}{Ni$_2$, $x^2-y^2$} &
   \multicolumn{1}{c}{Ni$_3$, $z^2$} & \multicolumn{1}{c}{Ni$_3$, $x^2-y^2$} &
   \multicolumn{1}{c}{Ni$_4$, $z^2$} & \multicolumn{1}{c}{Ni$_4$, $x^2-y^2$} \\
   1.73  & 1.07  & 0.32  & 0.31  & 0.63  & 0.46  & 0.39  & 0.44  \\
   1.07  & 1.88  & 0.31  & 0.29  & 0.45  & 0.37  & 0.43  & 0.50  \\
   0.32  & 0.31  & 1.73  & 1.07  & 0.39  & 0.42  & 0.63  & 0.46  \\
   0.31  & 0.29  & 1.07  & 1.88  & 0.44  & 0.50  & 0.45  & 0.37  \\
   0.63  & 0.45  & 0.39  & 0.44  & 1.82  & 1.12  & 0.32  & 0.31  \\
   0.46  & 0.37  & 0.42  & 0.50  & 1.12  & 1.89  & 0.31  & 0.29  \\
   0.39  & 0.43  & 0.63  & 0.45  & 0.32  & 0.31  & 1.82  & 1.12  \\
   0.44  & 0.50  & 0.46  & 0.37  & 0.31  & 0.29  & 1.12  & 1.89  \\
   \hline \\
   \multicolumn{8}{c}{$U_{mm'}^{\s\s}$ (parallel spin)} \\
   \multicolumn{1}{l}{Ni$_1$, $z^2$} & \multicolumn{1}{l}{Ni$_1$, $x^2-y^2$} &
   \multicolumn{1}{l}{Ni$_2$, $z^2$} & \multicolumn{1}{l}{Ni$_2$, $x^2-y^2$} &
   \multicolumn{1}{l}{Ni$_3$, $z^2$} & \multicolumn{1}{l}{Ni$_3$, $x^2-y^2$} &
   \multicolumn{1}{l}{Ni$_4$, $z^2$} & \multicolumn{1}{l}{Ni$_4$, $x^2-y^2$} \\
   0.00  & 0.74  & 0.32  & 0.31  & 0.60  & 0.45  & 0.38  & 0.43  \\
   0.74  & 0.00  & 0.31  & 0.29  & 0.45  & 0.37  & 0.42  & 0.48  \\
   0.32  & 0.31  & 0.00  & 0.74  & 0.38  & 0.42  & 0.60  & 0.45  \\
   0.31  & 0.29  & 0.74  & 0.00  & 0.43  & 0.48  & 0.45  & 0.37  \\
   0.60  & 0.45  & 0.38  & 0.43  & 0.00  & 0.83  & 0.32  & 0.31  \\
   0.45  & 0.37  & 0.42  & 0.48  & 0.83  & 0.00  & 0.31  & 0.29  \\
   0.38  & 0.42  & 0.60  & 0.45  & 0.32  & 0.31  & 0.00  & 0.82  \\
   0.43  & 0.48  & 0.45  & 0.37  & 0.31  & 0.29  & 0.82  & 0.00  \\
\end{tabular}
\end{table*}

\subsubsection{Orthorhombic \luno{}}

For the orthorhombic phase of \luno{}, the average density-density interaction
between electrons with opposite spins in the same orbital is found to be $U =
1.65$~eV, that between opposite spins in different orbitals is $U' = 0.99$~eV. 
The interaction between parallel spins in different orbitals is the smallest, 
in accordance with Hund's rule, being reduced to $U^{\sigma\sigma}=U'-J =
0.66$~eV.
We observe that, despite the orthorhombic distortion, these parameters obey almost perfectly 
the relation  $U' = U-2J$ expected for a cubic system, with $U = 1.65$~eV, $J = 0.33$~eV. 

The average nearest-neighbor parallel-spin interaction $V^{\s \s}_1$ is 
$0.42$~eV, where the average is taken evenly over both neighbors in the unit cell, 
comparable to the average on-site parallel-spin interaction $U' - J=0.66$~eV .
Additionally, the next-nearest-neighbor parallel-spin interaction $V^{\s
\s}_2$ can be estimated from e.g. Ni$_1$ and Ni$_2$ to be $0.30$~eV. 

\subsubsection{Monoclinic \luno{}}

For the monoclinic phase we obtain for the averaged parameters 
$U = 1.83$~eV, $U' = 1.09$~eV and $U'- J = 0.74$~eV, which is fairly
consistent with the Kanamori parametrisation for a cubic system with
$U = 1.83, J = 0.37$~eV's.

The on-site parallel-spin interaction $U^{\sigma\sigma}=U-3J = 0.74$~eV is again of
a similar order of magnitude to the average nearest-neighbor parallel-spin
interaction $V^{\s \s}_1 = 0.44$~eV. In this case, due to the distortions in
the structure, the average needs to be weighted to account for the
fact that Ni$_1$ has 4 Ni$_4$ atoms and 2 Ni$_3$ atoms as neighbors.
For the next-nearest-neighbor interaction, we obtain $V^{\s \s}_2 = 0.31$~eV.

\subsubsection{Long-range nature of interactions}

For both phases, we notice that the nearest-neighbor intersite interactions
$V_1$, for example between Ni$_1$ and Ni$_3$ and between Ni$_1$ and Ni$_4$, are
found to be non-negligible. 
It must be emphasized that the contributions to screening from particle-hole transitions within the 
$e_g$ manifold of bands are excluded in the cRPA procedure. Indeed, the effective interactions 
obtained from cRPA are to be used in the low-energy effective $e_g$ model, and further screening relies on the 
many-body treatment of this model (e.g. with DMFT). 
As a result, the interactions in the low-energy effective model 
are screened exclusively by interband transitions, as in an insulator, and thus one should 
expect significant long-range Coulomb interactions. Indeed, the second nearest-neighbor
interactions $V_2$ are likewise quite large.  Upon closer inspection, it is clear
that the interactions up to the second shell of neighbors 
decay as $1/R$, indicating that long-range interactions must be accounted for to reach an
accurate description of the physics of these materials.
The situation is comparable to the recently studied case of Sn/Si(111)
\cite{0953-8984-25-9-094005,PhysRevLett.110.166401, Hansmann_uncertainty}: 
there it was shown that the continuum limit that allows to
parametrize the interaction tail as $\frac{V_1}{R}$ with the nearest
neighbor interaction $V_1 = \epsilon^{-1} V_1^{bare}$, with $\epsilon$
the macroscopic dielectric constant, is already reached at nearest
neighbor distances. Similarly, in graphene, the long-range tail of the
interactions was argued to be responsible for the necessary screening
to prevent the system from becoming a Mott insulator 
\cite{PhysRevLett.111.036601}.
\comment{
More generally, the effects of non-local interactions in correlated 
materials and models thereof have recently raised tremendous interest
in the community \cite{PhysRevLett.87.237002,PhysRevB.82.155102,ayral_gwdmft,ayralPRB87-13,PhysRevB.90.195114,PhysRevB.90.245115,
PhysRevB.92.081106, 
PhysRevB.94.239906,PhysRevB.94.165141,PhysRevB.95.245130,PhysRevB.95.115149}, 
within different lattice geometries.
}

From Tables \ref{tab:u_orthorhombic} and \ref{tab:u_monoclinic} one may also
conclude that the spin dependence of intersite interactions is negligible; the
exchange interaction arises due to direct overlap of the $e_g$ orbitals and,
therefore, is well localized. Hence, from now on we will suppress the spin
subscripts in the intersite interactions $V$. 

%
%
\section{Physical consequences for \luno{} and DMFT calculations}
\label{sec:longrange_dmft}

\subsection{Effective theory for low-energy $e_g$ states}

From the sections above, we can infer the following effective Hamiltonian for a description of 
\luno{} involving only low-energy $e_g$ states: 
\begin{equation}
\hH\,=\,\hH_0 + \sum_i \hH^{(i)}_U+\hH_V.
\label{eq:ham}
\end{equation}
In this expression, $\hH_0$ is the single-electron part of the effective Hamiltonian.
Within the DFT+DMFT framework, $\hH_0$ is constructed as:
\begin{equation}
\hH_0\,=\,\hH^0_{\mathrm{DFT}}- \hH_{dc},
\label{eq:H0}
\end{equation}
with $\hH^0_{\mathrm{DFT}}$ the single-electron Kohn-Sham Hamiltonian for 
$e_g$ bands, as obtained from DFT(-LDA) and $\hH_{dc}$ is a double-counting correction to be detailed below. 
The many-body terms $\hH_U$ and $\hH_V$ are local (on-site) and inter-site interaction terms, respectively. 
For the local term, the full Kanamori Hamiltonian appropriate for $e_g$ states is considered, 
namely on each lattice site $(i)$: 
\begin{align*}
\hH_U = & U  \sum_{m} \hn_{m \up} \hn_{m \dn}
              + (U - 2J) \sum_{m \neq m'} \hn_{m \up} \hn_{m' \dn} \\
            & + (U - 3J) \sum_{m < m', \s} \hn_{m \s} \hn_{m' \s} \\
            & - J \sum_{m \neq m'} c^\dagger_{m \up} c_{m \dn} c^\dagger_{m' \dn} c_{m' \up}
              + J \sum_{m \neq m'} c^\dagger_{m \up} c^\dagger_{m \dn} c_{m' \dn} c_{m' \up}.
\end{align*}
The inter-site term is taken to be of the form: 
\begin{equation*}
\hH_V\,=\,\frac{1}{2} \sum_{i\neq j} V_{ij} \hn_i \hn_j.
\end{equation*}
In this expression, the coupling constants $U$, $J$ and $V_{ij}=V_1/R_{ij}$  
(with $R_{ij}$ the distance between the two atomic sites $i,j$) 
are determined from the cRPA calculations presented above. 

In the following, we show that this low-energy effective model with cRPA values of the interaction parameters 
provides a satisfactory description of the physics of \luno{}. This is done by using the DFT+DMFT framework in 
order to construct and solve the low-energy model. 
We find that intersite interactions must be taken into account in this low-energy description: 
they are included in our calculations at the level of Hartree mean-field theory. 
Finally, these findings are discussed in relation to the low-energy picture of rare-earth nickelates proposed 
in Ref.~\onlinecite{subedi_low-energy_2015}.

\subsection{Hartree treatment of long-range interactions}
\label{sec:madelung}

In order to take the intersite terms into account we employ the
Hartree approximation for long-range interactions. Note 
that this is consistent with the DMFT approach, in which only local interactions 
have dynamical effects, while non-local interactions are treated at the 
static mean-field level. 
In the Hartree approximation, $\hH_V$ reduces to:
\begin{equation*}
\hH_V \rightarrow \frac{1}{2} \sum_{i\neq j} V_{ij} \left[n_i \hn_j+n_j \hn_i - n_i n_j \right],
\end{equation*}
with the effective Hartree one-body Hamiltonian and potential
\begin{equation*}
 \hH_{\rm eff} = \sum_i V_H(i) \hn_i\,\,\, , \,\,\, V_H(i) = \sum_{j \neq i} V_{ij} n_j.
\end{equation*}

The total energy from the interacting part of the Hamiltonian is thus
\begin{equation}
E_V[\{n_i\}]\,=\,\frac{1}{2} \sum_{i\neq j} V_{ij} n_i n_j = \frac{1}{2} \sum_i V_H(i) n_i.
\label{eq:en_v}
\end{equation}

Let us now consider the present case of the Ni sublattice in  \luno{}. It can be well approximated by the NaCl-type bipartite lattice
with two inequivalent, LB and SB, sublattices with occupancies $n_{\LB}$ and $n_{\SB}$ per site, respectively.
Given that the system is charge neutral the formally diverging term on the r.h.s. of Eq.~\eqref{eq:en_v}
can be summed using the Madelung method, resulting in the following sublattice potentials:
\begin{align*}
V_H(\LB) = & - M V_1 (n_{\LB}-n_{\SB})/2, \\
V_H(\SB) = & + M V_1 (n_{\LB}-n_{\SB})/2,
\end{align*}
where $M$ is the Madelung constant for the NaCl lattice and the uniform part
of the potential is dropped.

By comparing this result with the Hartree term with only nearest-neighbor
interactions,
\begin{align*}
V_H^{nn}(\LB) = & - z_{\rm eff} V_1\,(n_{\LB}-n_{\SB})/2, \\
V_H^{nn}(\SB) = & + z_{\rm eff} V_1\,(n_{\LB}-n_{\SB})/2,
\end{align*}
we can identify $M$ as an effective connectivity of the lattice
\begin{equation*}
z_{\mathrm{eff}} = M \approx 1.747 .
\end{equation*}
Note that the effect of the Madelung summation is that the effective connectivity $z_{\mathrm{eff}}$ is 
significantly reduced as compared to the lattice connectivity $z=6$.

The Hartree potential above can be viewed as a site-dependent contribution to the 
self-energy coming from the intersite interactions, which reads: 
\begin{equation}
\Sigma^{V}_{\alpha} = z_{\rm eff} V_{1} (n_{\bar{\alpha}}-n_{\alpha})/2.
\end{equation}
In this expression, $\bar{\alpha}$ designates the opposite sublattice relative to $\alpha$,
i.e. if $\alpha=\SB$ then $\bar{\alpha}=\LB$ and vice versa; $n_{\alpha}$ is the
$e_g$ occupancy of the corresponding site. 

\subsection{Atomic limit}
\label{sec:atomic_limit}

Before discussing the DFT+DMFT results, we first consider the atomic limit in which 
all hopping terms in the Hamiltonian (\ref{eq:ham}) are set to zero so that $\hH_0$ contains 
only an on-site Peierls potential equal to $-\Delta_s/2$ on LB sites and $+\Delta_s/2$ on SB sites
(see Fig.~\ref{fig:disproportionation}). 
We will compare in this limit the energies of two states: the uniform one (UN) $n_{LB}=n_{SB}=1$ and 
the fully disproportionated one (FD) $n_{LB}=2, n_{SB}=0$. 

The contribution of the on-site Peierls potential to the energy is 
$-\Delta_s\sum_{i\in LB}n_i/2 +\Delta_s\sum_{i\in SB}n_i/2$: it vanishes in the uniform state 
and provides an energy gain $- \Delta_s (N_s/2)$ in the FD state, with $N_s$ the total number of 
lattice sites. The on-site interaction energy vanishes too in the UN state, and is equal to 
$+U^{\s\s} N_s/2$ in the FD state with $U^{\s\s}=U-3J$, since two electrons on a LB site will occupy the 
high-spin Hund's rule configuration with one electron in each of the two $e_g$ orbitals. 
Finally, using the above expressions in the Hartree approximation, the contribution of the 
inter-site interactions to the energy reads:
\begin{equation}
\langle \hH_V\rangle = -\frac{N_s}{8} z_{\rm eff} V_1 (n_{\LB}-n_{\SB})^2.
\end{equation}
It vanishes again in the UN state, and provides an energy gain $-V_1 z_{\rm eff} N_s/2$ in the FD state. 
Hence, the energy difference between the FD state and the uniform one reads, in the atomic limit: 
\begin{eqnarray}\nonumber
E_{FD} - E_{UN}&=&\frac{N_s}{2} \left[U^{\s\s}- z_{\rm eff} V_1 -\Delta_s\right]\\ 
&=&\frac{N_s}{2} \left[U-3J - z_{\rm eff} V_1 -\Delta_s\right]
\end{eqnarray}

The transition into the
charge-disproportionated state in the atomic limit occurs, therefore, when 
\begin{equation}
U^{\s\s} - z_{\rm eff} V_1 < \Delta_s.
\label{eq:inst} 
\end{equation}
Note that, if only the nearest-neighbor component of the non-local interactions is taken 
into account, the FD state is stable for $U^{\s\s} - z V_1 < \Delta_s$. The above criterion 
in the presence of long-range interaction simply amounts to replacing the connectivity of the lattice 
$z$ by the effective Madelung connectivity. 

Let us consider the orthorhombic phase where $\Delta_s=0$. The above criterion then reads 
$U-3J-z_{\rm eff} V_1 < 0$. Hence, a small enough value of $U$ (e.g. strongly reduced 
by screening) or a large enough value of the Hund's coupling $J$ leads to an instability 
into the disproportionated state, as noted in previous work~\citep{mazin_charge_2007,subedi_low-energy_2015}. 
In the present context, this instability is a spontaneous symmetry breaking of 
electronic origin, since there is only one type of sites in this crystal structure. 
Our cRPA results for \luno{} in the orthorhombic phase yield 
$U^{\s \s} - z_{\rm eff} V_1 \simeq -0.1 eV$: the combined effect of 
screening and long-range interactions yields a small but negative value of this  
quantity, which is consistent with the physical picture of Subedi et al.\cite{subedi_low-energy_2015}. 
Hence, in the atomic limit, we would conclude that the orthorhombic phase is spontaneously 
unstable to disproportionation! In reality, as shown below, the inclusion of inter-site 
hopping in a full DFT+DMFT treatment leads to the correct conclusion that the orthorhombic phase 
is not electronically disproportionated - the atomic-limit estimate providing a considerable 
overestimation of the range of stability of the FD state. However, inaccurate as it may 
be (especially in the metallic state), the virtue of this atomic limit estimate is 
to emphasize how screening, a large $J$, and sizeable inter-site interactions can lead 
to disproportionation. 

In the monoclinic phase, we obtained $U^{\s\s} = 0.74$~eV, $V_1 = 0.44$~eV, hence 
$U^{\s \s} - z_{\rm eff} V_1 \simeq -0.03$~eV. Basically any positive value of the 
Peierls energy gap, which is non-zero in this phase, will 
thus stabilize a fully disproportionated state in the atomic limit. As shown below, 
monoclinic \luno{} is indeed found to be a disproportionated insulator when performing DFT+DMFT 
calculations with these interaction parameters.  

\subsection{DMFT: setup and double counting}

We now turn to the results obtained in the DFT+DMFT framework, providing first 
some technical details about the calculation. 

The one-electron part of the  effective Hamiltonian is $\hH_0\,=\,\hH^0_{\mathrm{DFT}}- \hH_{dc}$. 
The DFT Hamiltonian was obtained using the FLAPW method as implemented in the Wien2k software package \cite{wien2k},
with Perdew-Burke-Erzenhof (PBE) approximation \cite{pbe96} for the exchange-correlation functional. 
A $\mathbf{k}$-mesh of $6\times5\times4$ points is used. 
Projected local orbitals\cite{Amadon2008} spanning the low-energy $e_g$ subspace are constructed using the implementation 
of the \textsc{TRIQS/DFTTools} software package
\cite{Aichhorn2009,aichhorn_dfttools_2016,parcollet_triqs_2015,triqs}. 

The full local self-energy arises from both the DMFT treatment of the local interactions $H_U$ 
in (\ref{eq:ham}) at the dynamical level and from the non-local interactions treated within the 
Hartree approximation, namely: 
\begin{equation}
\Sigma_{\alpha}(i\omega_{n}) = \Sigma^{\mathrm{imp}}_{\alpha}(i\omega_{n})-\Sigma^{\mathrm{imp}}_{dc,\alpha}\,
+\,\Sigma^{V}_{\alpha}-\Sigma^{V}_{dc,\alpha}.
\end{equation}
In this expression, $\alpha=\mathrm{LB,SB}$ is an index labelling LB and SB sites, 
$\Sigma^{V}_{\alpha}$ is the Hartree self-energy: 
\begin{equation}
\Sigma^{V}_{\alpha} = z_{\rm eff} V_{1} (n_{\bar{\alpha}}-n_{\alpha})/2,
\end{equation}
and $\Sigma^{\mathrm{imp}}_{\alpha}(i\omega_{n})$ is obtained by solving 
the DMFT effective impurity model using the
hybridization-expansion continuous-time quantum Monte Carlo (CTQMC) algorithm
\textsc{TRIQS/cthyb} \cite{seth_cthyb_2016,parcollet_triqs_2015,triqs}.   

A double-counting (DC) correction must be included, in order to remove the contribution 
from interactions already included within DFT. This DC correction can be viewed equivalently 
as the $\hH_{dc}$ part of $\hH_0$ or as part of the self-energy. 
The DC correction to the self-energy arising from the $U,J$ interactions is evaluated 
in the fully-localized limit\cite{Anisimov1997} as follows \cite{Aichhorn2009}:
\begin{equation}\label{eq:dc1}
\Sigma^{\mathrm{imp}}_{dc,\alpha}=\bar{U}(n_{\alpha}^{\DFT}-1/2)-\bar{J}(n_{\alpha}^{\DFT}/2-1/2),
\end{equation}
while the DC correction to the Hartree self-energy associated with the long-range 
interactions reads:
\begin{equation}
\Sigma_{dc,\alpha}^{V}=z_{\rm eff} V_{1} (n^{\DFT}_{\bar{\alpha}}-n^{\DFT}_{\alpha})/2.
\end{equation} 
In these expressions, $\bar{U}=U-J$ is the average interaction between electrons with opposite
spins, $\bar{J}=\bar{U}-U^{\sigma\sigma}=2J$, and $n_{\alpha}^{\DFT}$ is the
occupancy of Ni $e_g$ shell  in DFT for the site $\alpha=\LB$ or $\SB$. 

In the monoclinic phase the two inequivalent Ni sites have different $e_g$ occupancy
already at the DFT level: $n_{\alpha}^{\DFT}=1.17$ and $0.83$ for the LB and SB site, 
respectively. From the expressions above, one sees that the Peierls energy splitting 
between LB and SB sites appearing in $\hH_0\,=\,\hH^0_{\mathrm{DFT}}- \hH_{dc}$
is renormalized by double counting, as compared to its DFT value: 
\begin{equation}
\Delta_s\,=\,\Delta_s^{\DFT} + \left(U-2J-z_{\rm eff} V_1\right) \Delta n^{\DFT},
\end{equation}
with $\Delta n^{\DFT}=n_{\LB}^{\DFT}-n_{\SB}^{\DFT}$. Note that given the cRPA values above, 
$U-2J-z_{\rm eff} V_1\simeq 0.32$~eV is positive so that double-counting enhances the 
effective value of the Peierls energy, from $\simeq 0.25$~eV at the DFT level to 
$\Delta_s\simeq 0.36$~eV. 

\subsection{DMFT: results and phase diagram}
\label{sec:phase_diagram}

\begin{figure}[htpb]
  \centering
    \includegraphics[width=1.00\columnwidth]{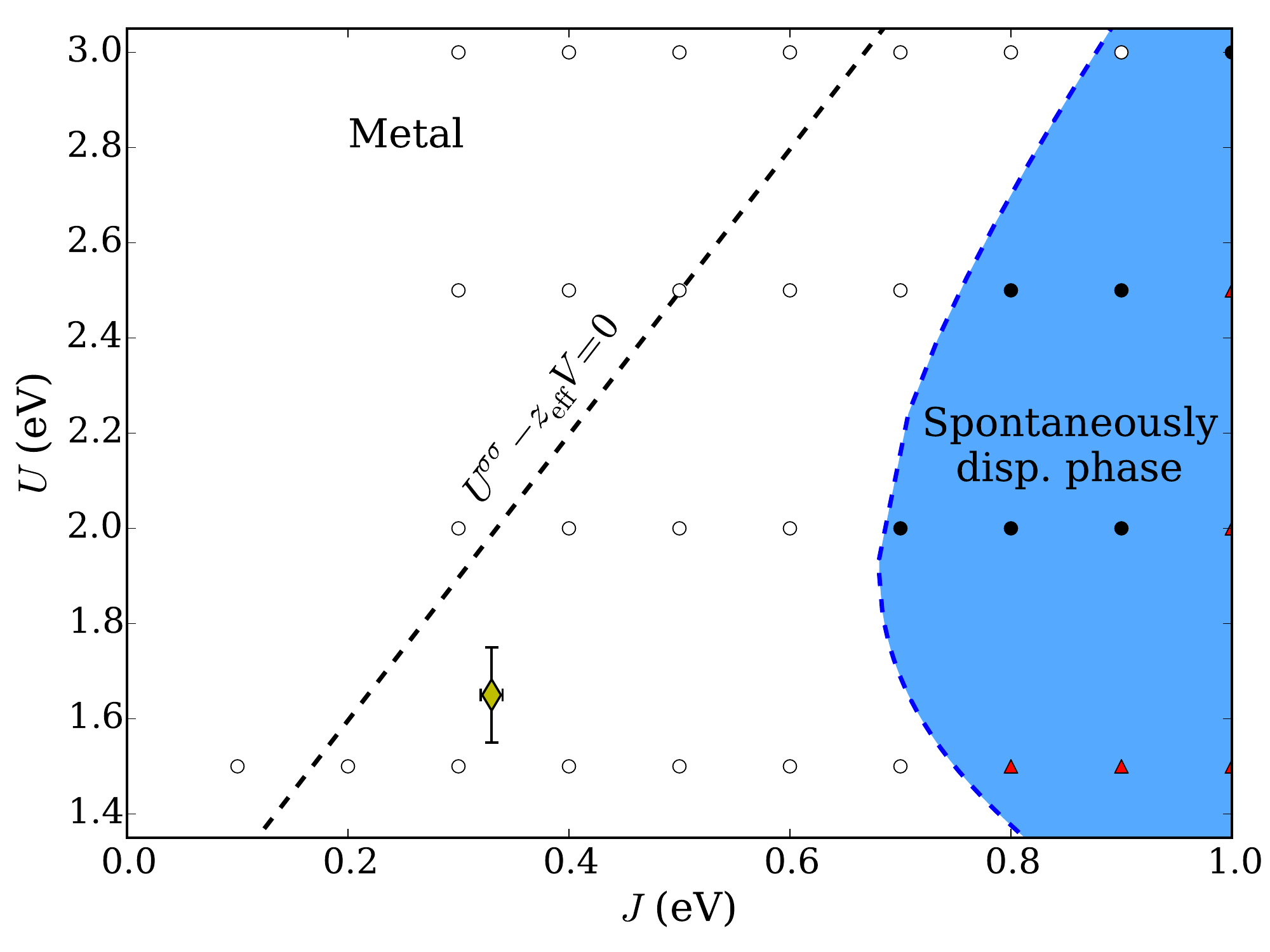}

    \includegraphics[width=1.00\columnwidth]{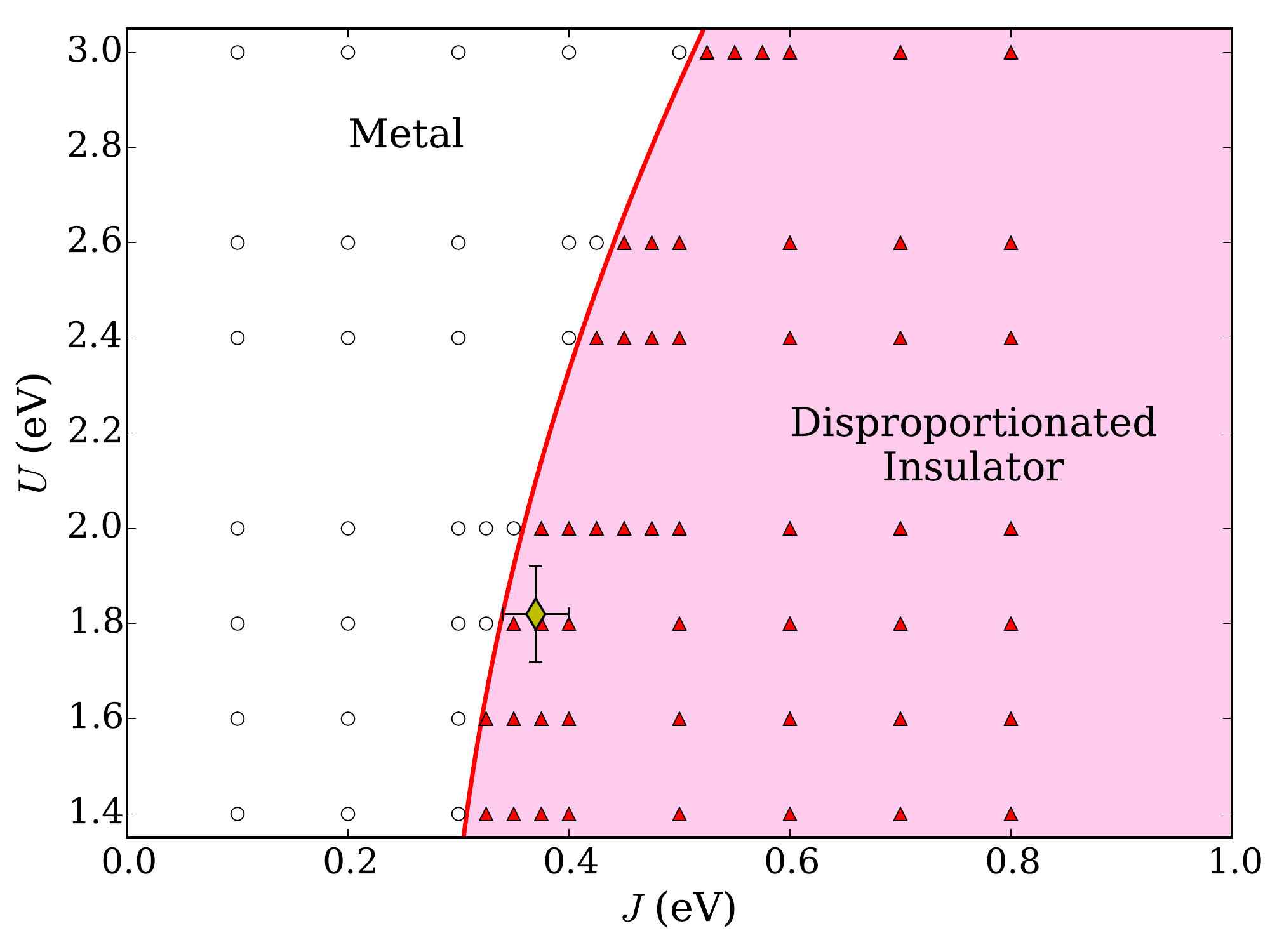}
        \caption{Phase diagrams of the monoclinic (bottom) and orthorhombic
(top) phases of \luno{}. For the monoclinic phase: empty circles show metallic
solutions, red triangles correspond to a disproportionated insulator.
For the orthorhombic phase: empty circles stand for uniform metallic,
filled circles -- for disproportionated metallic, and red triangles -- for
disproportionated insulating solutions; the blue domain contains spontaneously
disproportionated (metallic or insulating) solutions. The value of $V_{1}$ is fixed to the average cRPA
value.  The cRPA values of $U$ and $J$ are marked by the diamond, with error
bars showing variations for different sites and orbitals. 
 \label{fig:phase_diag}}
\end{figure}

In order to explore how the values of the interaction strengths affect the 
physics of \luno{} in each crystal structure, we have performed a series of DFT+DMFT 
calculations for a fixed value of $V_{1} = 0.44$~eV with varying $U$ and $J$.
The obtained phase diagrams are presented in Fig.~\ref{fig:phase_diag}.
The main qualitative features are similar to the results of 
Ref.\onlinecite{subedi_low-energy_2015}, in which the non-local interactions were not 
taken into account and only the local Kanamori interactions were included. 
Specifically, both the monoclinic and orthorhombic
structures have a phase boundary separating a uniform metallic 
and a disproportionated (insulating or metallic) phase.

We note that the location of this boundary is very different for the orthorhombic 
and for the monoclinic phase, being pushed towards much smaller values of $J$ for the 
latter. This demonstrates the strong sensitivity of the disproportionation to the 
value of the Peierls energy~\cite{subedi_low-energy_2015}. 

The physical range of interaction parameters must be associated with regions of 
the phase diagram corresponding to the monoclinic phase being insulating
and the orthorhombic one being metallic with uniform distribution of site occupancies
(non-disproportionated metal). Because of the great sensitivity of the critical boundary to 
$\Delta_{s}$, rather extended regions of the $(U,J)$ parameter space satisfy these conditions 
(basically corresponding to the area delimited by the orthorhombic (blue) boundary to the right 
and the monoclinic (red) one to the left, including the range
$J = 0.3-0.7$ and $U = 1.5-2.0$). 

The calculated cRPA values are marked by a (yellow) diamond symbol on each panel of 
Fig.~\ref{fig:phase_diag}. 
They are located well within the metallic domain for the orthorhombic phase and
just inside the insulating domain (rather close to the MIT boundary) for the monoclinic phase. 
These results demonstrate that cRPA is able to provide reasonable values of the effective screened 
interactions, which correctly account for the physical nature of each phase. 

\section{Discussion and conclusions}
\label{sec:discussion}

In summary, the question we have addressed in this article is that of the 
appropriate values of interaction parameters for rare-earth nickelates, when 
adopting a low-energy description of their electronic structure involving 
only the partially occupied $e_g$ states. 
We have calculated these effective low-energy interaction parameters from first principles for \luno{}, 
using the constrained random-phase approximation (cRPA). 
The obtained values confirm the strong reduction of the effective on-site $U$ by screening, 
down to $U\simeq 1.65$~eV in the orthorhombic phase ($U\simeq 1.83$~eV in the monoclinic phase), 
while the Hund's coupling $J$ remains sizeable ($J\simeq 0.33, 0.37$~eV in each phase, 
respectively).

The cRPA results also reveal the importance of the long-range intersite interactions, 
with a slow spatial decay $V_1/R$. $V_1$ is found to be of order $0.42-0.44$~eV so that these 
interactions must be included in a proper low-energy treatment. 
When treated at the level of a Hartree approximation, they lead to a further 
reduction of the effective parallel-spin local interaction 
$\Ueffuu = U - 3J - z_{\rm eff}V_{1}$ (with $z_{\rm eff}$ the effective Madelung connectivity), 
which is found to be small and negative. 
This is qualitatively consistent with the picture of a negative charge-transfer 
insulator and validates the low-energy description advocated in earlier 
work\cite{mazin_charge_2007,subedi_low-energy_2015}. 
Let us also note that the low-energy interaction is further renormalized by higher-order many-body effects
not taken into account in the present work, as well as by dynamical screening due to phonon modes. 
A rough estimate of the latter effect in nickelates suggests that it is small, 
but a more detailed study is left for future work. 
We have constructed an appropriate low-energy model based on the cRPA effective interactions, 
and solved this model in the DFT+DMFT framework. We found that the monoclinic structure falls 
within the bond-disproportionated-insulator region, while the orthorhombic
structure is located deep in the uniform metallic state, in agreement with experimental 
observations.
While our calculations take into account only electronic degrees of freedom, 
a full theory of the metal-insulator transition in nickelates should also take into account 
the coupling to the relevant lattice mode associated with the structural transition: 
this should be addressed in future work.

Besides the specific example of nickelates, our work can be put in the 
broader context of compounds with small or negative charge transfer leading to the 
possible formation of ligand holes. 
We have shown that it is possible to build an appropriate low-energy effective theory of 
such compounds, involving only electronic states near the Fermi level, provided the 
strong reduction of the low-energy effective interactions is properly taken into account. 
This provides a perspective on these materials which is complementary to the 
one in which ligand states are explicitly retained in the 
description~\cite{Mizokawa1991,Mizokawa1994,Mizokawa2000}. 
Future work should document the general applicability of the present approach 
by considering other compounds with small or negative charge transfer.

\begin{acknowledgments}
We are grateful to Manuel Bibes, Sara Catalano, Claude Ederer, Marta Gibert,
Alexander Hampel, Marisa Medarde, Andrew J. Millis, Yusuke Nomura, Swarup
Panda, Julien Ruppen, George Sawatzky, Hugo Strand, Alaska Subedi, Jean-Marie
Tarascon, J\'er\'emie Teyssier, Jean-Marc Triscone, Dirk van der Marel, and
Julien Varignon for numerous discussions about the physics of rare-earth
nickelates.  
This work was supported by the European Research Council grants ERC-319286-`QMAC' (A.Georges), 
ERC-278472-`MottMetals' (O.Parcollet, P.~S.), ERC-617196-`CorrelMat' (S.Biermann, P.~S.) 
and IDRIS/GENCI Orsay under project 201601393.
L.~Pourovskii acknowledges computational resources provided by the
Swedish National Infrastructure for Computing (SNIC) at the
National Supercomputer Centre (NSC) and PDC Center for
High Performance Computing.
O.~Peil and A.~G. acknowledge support from the Swiss National Science Foundation NCCR MARVEL and 
computing resources provided by the Swiss National Supercomputing Centre (CSCS) under projects s575 
and mr17.
M. B. gratefully acknowledges financial support from the Helmholtz Association through 
the Hemholtz Postdoctoral Programme (VH-PD-022).
F. A. acknowledges financial support from the Swedish Research Council (VR).
\end{acknowledgments}

\bibliography{main}

\end{document}